\documentclass[aps,prl,reprint,superscriptaddress,nofootinbib]{revtex4-2}

\usepackage[T1]{fontenc}
\usepackage[utf8]{inputenc}
\usepackage{amsmath,amssymb,bm}
\usepackage{graphicx}
\usepackage{hyperref}

\usepackage{tikz}
\usepackage{pgfplots}
\pgfplotsset{compat=1.18}
\usetikzlibrary{arrows.meta,positioning,fit,calc}

\begin{document}

\title{Variational Boundary Fluctuations as a First-Principles Origin of Langevin Noise}

\author{Francisco Monroy}
\email{monroy@ucm.es}
\affiliation{Departamento de Química Física, Universidad Complutense de Madrid, 28040 Madrid, Spain}

\date{\today}

\begin{abstract}
Stochastic forces are usually postulated or obtained by eliminating
environmental degrees of freedom. Here we identify a variational origin:
fluctuating endpoint data in Hamilton's principle induce fluctuations of
the on-shell action. Hamilton--Jacobi propagation transports this
boundary imprint, whose gradient generates an effective Langevin force
inherited from boundary-action fluctuations. The resulting force is not
freely specifiable: its amplitude is filtered by the Hessian of
Hamilton's principal function, yielding multiplicative and
state-dependent noise. Homogeneous additive Langevin forcing is recovered
only as a Markovian coarse-grained limit.
\end{abstract}

\maketitle

\begin{figure}[htb!]
\centering
\includegraphics[width=\columnwidth]{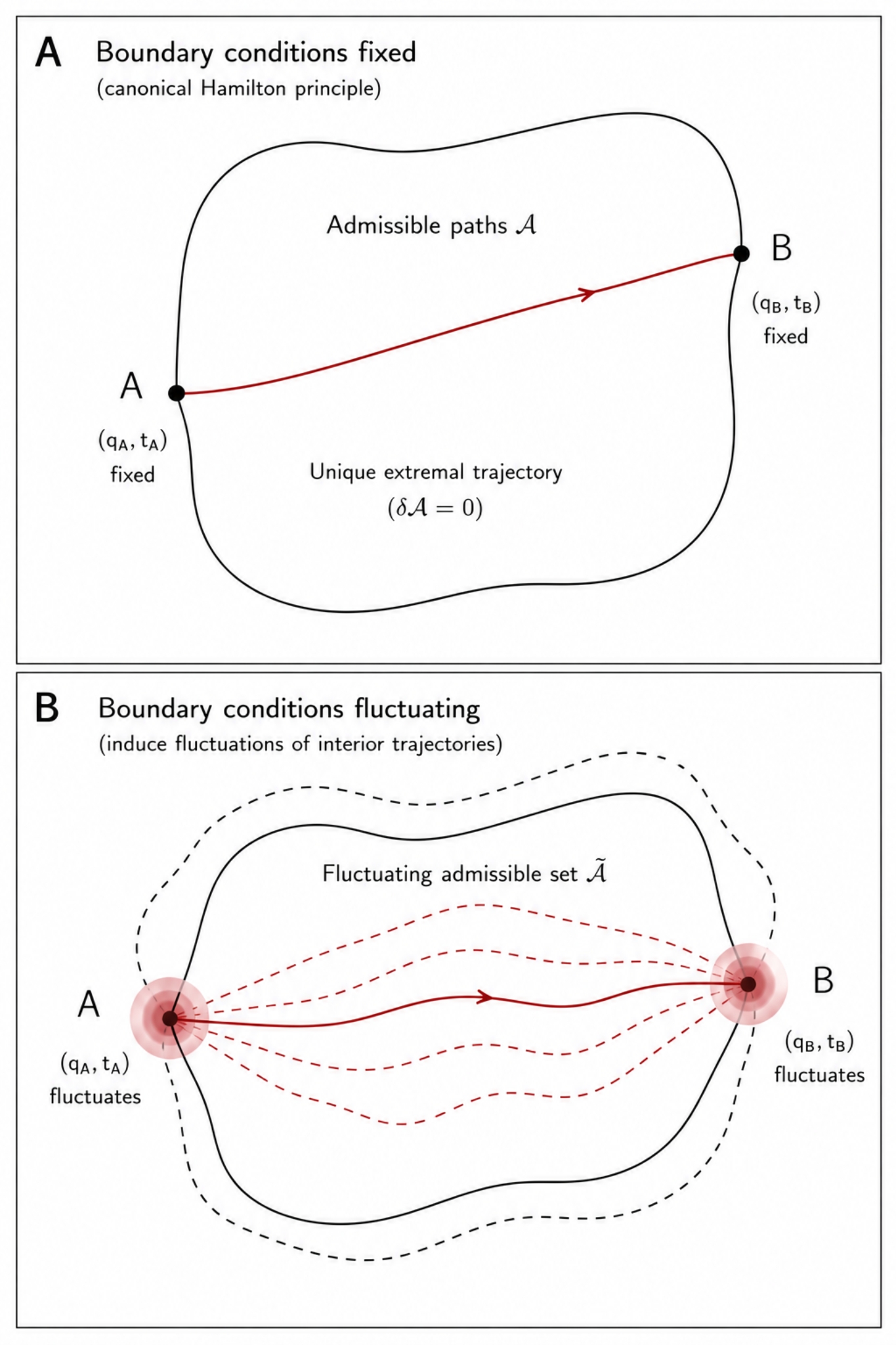}
\caption{
\textbf{Variational origin of boundary-induced trajectory fluctuations.}
(A) With fixed endpoints, the admissible variational set
\(\mathcal A\) selects a reference extremal trajectory between states
\(A\) and \(B\), satisfying \(\delta\mathcal A=0\).
(B) Environmental uncertainty in preparation and termination makes
endpoint constraints fluctuate, shown by red halos
\(\eta_i,\eta_f\). This selects neighbouring admissible shells,
schematically represented by \(\widetilde{\mathcal A}\), and generates a
family of nearby extremals, shown as dashed trajectories. The bulk
dynamics remains deterministic for each realization: endpoint
fluctuations induce a boundary variation of Hamilton's principal function,
which is then propagated by the Hamilton--Jacobi flow.
}
\label{fig:boundary_fluctuations}
\end{figure}

Hamilton's principle is normally formulated with fixed endpoints. For a
trajectory \(q(t)\), the Lagrangian action
\begin{equation}
\mathcal A[q]
=
\int_{t_i}^{t_f}
L(q,\dot q,t)\,dt
\label{eq:intro_action}
\end{equation}
has first variation
\begin{equation}
\delta\mathcal A
=
\int_{t_i}^{t_f}
\left(
\frac{\partial L}{\partial q}
-
\frac{d}{dt}
\frac{\partial L}{\partial\dot q}
\right)\delta q\,dt
+
\left[p\,\delta q\right]_{t_i}^{t_f},
\label{eq:intro_variation}
\end{equation}
with \(p=\partial L/\partial\dot q\). The standard choice
\(\delta q(t_i)=\delta q(t_f)=0\) removes the endpoint term and yields
the Euler--Lagrange equations \cite{Landau1976}, defining the
fixed-boundary basis of analytical mechanics
\cite{Goldstein2002,Arnold1989}. Physical acts, however, need not have
perfectly sharp preparation and termination constraints
\cite{Brillouin1962}. We therefore consider a nearby family of
variational problems with fluctuating endpoint data and ask how the
associated on-shell boundary-action perturbation propagates into the
dynamics. As illustrated in Fig.~\ref{fig:boundary_fluctuations}, fixed
endpoints select a reference extremal, whereas environmental uncertainty
selects neighbouring extremals within a nearby boundary-value family.

Stochastic dynamics is usually obtained by other routes. Langevin theory
postulates fluctuating forces directly in the equations of motion
\cite{Langevin1908}. Projection-operator methods and generalized
Langevin equations generate effective noise by eliminating environmental
degrees of freedom \cite{Mori1965,Kubo1966,Zwanzig2001}. Influence-functional
and Brownian-bath models produce fluctuations through hidden reservoirs
\cite{Feynman1963,Ford1965,Caldeira1983}. Related stochastic structures
also appear in stochastic variational mechanics \cite{Nelson1966,Yasue1981},
stochastic control and Hamilton--Jacobi--Bellman theory
\cite{Fleming2006,LazaroCami2009}, fluctuating interfaces
\cite{Kardar1986}, and stochastic partial differential equations with
boundary-driven forcing \cite{Walsh1986,DaPrato1992,Flandoli1995}. In
these frameworks, randomness enters the bulk dynamics, a bath, or an
underlying stochastic evolution law. Here the Hamiltonian bulk remains
deterministic, while uncertainty enters only through the variational
boundary data.

On shell, endpoint fluctuations perturb Hamilton's principal function
through the boundary action variation
\(\delta S_\partial=p_f\eta_f-p_i\eta_i\), where
\(\eta_i,\eta_f\) are endpoint displacements. This perturbation is then
transported by the Hamilton--Jacobi flow. Its local residual under the
reference transport defines an effective action source,
\(\zeta=-D_t^{(0)}S_\partial\); taking the gradient of this source yields
an effective force, \(\xi=\nabla\zeta\). Thus the causal object is not a
primitive Langevin force, but a transported perturbation of the action
field. For mechanical Hamiltonians, the inherited force amplitude is
filtered by the Hessian \(\nabla^2S\), giving multiplicative,
anisotropic, and state-dependent noise. Homogeneous additive Langevin
forcing appears only after Markovian coarse graining.

The mechanism developed below is
\begin{equation}
\delta q_\partial
\;\Rightarrow\;
\delta S_\partial
\;\Rightarrow\;
\zeta=-D_t^{(0)}S_\partial
\;\Rightarrow\;
\xi=\nabla\zeta .
\label{eq:intro_chain}
\end{equation}
Thus stochastic forcing is not assumed at the level of trajectories:
it is obtained by projecting an ensemble of deterministic extremals
through Hamilton--Jacobi propagation. We now derive this chain explicitly.

\paragraph*{Variational boundary fluctuations.}

Along each extremal trajectory in the boundary-value family, the bulk
variation vanishes and the on-shell action obeys the exact endpoint
differential
\begin{equation}
dS
=
\left[
p\,dq
-
H\,dt
\right]_{t_i}^{t_f},
\qquad
H=p\dot q-L .
\label{eq:onshell_differential}
\end{equation}
For fixed initial and final times, fluctuating endpoint displacements
\(\eta_i=\delta q_i\) and \(\eta_f=\delta q_f\) give
\begin{equation}
\delta S_\partial
=
p_f\eta_f-p_i\eta_i .
\label{eq:boundary_fluctuation}
\end{equation}
Thus the principal action acquires a boundary-induced perturbation. To
first order in the endpoint displacement,
\begin{equation}
S_\eta
=
S_0
+
S_\partial
+
O(\eta^2),
\qquad
S_\partial\equiv\delta S_\partial .
\label{eq:perturbed_principal_action}
\end{equation}
Here \(S_\eta\) labels the on-shell action selected by displaced boundary
data, not a stochastic modification of the Lagrangian. The bulk dynamics
remains deterministic for each realization: boundary uncertainty perturbs
the variational admissibility, and hence the principal action, before any
effective force appears. The full endpoint differential, including
fluctuating times, second-order corrections, and two-boundary
correlations, is given in the Supplemental Material.

\paragraph*{Hamilton--Jacobi propagation.}

We regard the principal function as a field over the final configuration
point, \(S(q,t)\equiv S(q,t;q_i,t_i)\), and write locally
\(S(q,t)=S_0(q,t)+S_\partial(q,t)\), where \(S_\partial\) is the
transported boundary imprint. This split assumes a single-valued
Hamilton--Jacobi branch. The unperturbed action satisfies
\begin{equation}
\partial_tS_0+H(q,\nabla S_0,t)=0 .
\label{eq:HJ_unperturbed}
\end{equation}
Substitution into the Hamilton--Jacobi equation uses the first-order
expansion
\begin{align}
H(q,\nabla S_0+\nabla S_\partial,t)
&=
H(q,\nabla S_0,t)
\nonumber\\
&\quad+
\frac{\partial H}{\partial p}
\bigg|_{p=\nabla S_0}
\cdot
\nabla S_\partial
+
O(S_\partial^2).
\label{eq:H_expansion_main}
\end{align}
Keeping the linear order gives the boundary-transport balance
\begin{equation}
\left(
\partial_t
+
v_0\cdot\nabla
\right)S_\partial
=
-\zeta(q,t),
\quad
v_0(q,t)
=
\frac{\partial H}{\partial p}
\bigg|_{p=\nabla S_0}.
\label{eq:boundary_transport_HJ}
\end{equation}
Equivalently,
\(\zeta=-D_t^{(0)}S_\partial\), with
\(D_t^{(0)}=\partial_t+v_0\cdot\nabla\). If the boundary imprint is only
advected by the reference flow, then \(\zeta=0\). For fluctuating
boundary data, \(\zeta\) measures the local failure of this free
advection: it is the residual of the boundary-action imprint under
Hamilton--Jacobi transport, not a primitive stochastic force inserted in
the bulk. The same residual may be represented as an effective source acting on
the full principal function,
\begin{equation}
\partial_tS+H(q,\nabla S,t)=\zeta(q,t).
\label{eq:HJ_source_representation}
\end{equation}
Below we use the full material derivative \(D_t=\partial_t+v\cdot\nabla\),
with \(v=\nabla S/m\); its difference from \(D_t^{(0)}\) is beyond the
retained linear order. Details are given in the Supplemental Material.

\paragraph*{Emergent force and geometric filtering.}

For a mechanical Hamiltonian \(H(q,p)=p^2/2m+V(q)\), with
\(p=\nabla S\) and \(v=p/m\), taking the gradient of
Eq.~\eqref{eq:HJ_source_representation} gives
\begin{equation}
\partial_t p+(v\cdot\nabla)p
=
-\nabla V+\nabla\zeta .
\label{eq:momentum_transport_main}
\end{equation}
Along a Hamilton--Jacobi characteristic,
\(D_t p=\partial_t p+v\cdot\nabla p=m\ddot q\), and therefore
\begin{equation}
m\ddot q
=
-\nabla V
+
\xi(q,t),
\qquad
\xi=\nabla\zeta .
\label{eq:effective_Langevin}
\end{equation}

Thus the effective Langevin force is the gradient of a transported action
residual. Its structure is fixed by the endpoint origin of
\(S_\partial\). In a neighbourhood where the Hamilton--Jacobi
characteristics remain single-valued, an endpoint displacement gives the
local first-order imprint
\begin{equation}
S_\partial(q,t)
\simeq
\nabla S(q,t)\cdot\eta(t),
\label{eq:local_boundary_structure}
\end{equation}
where \(\eta(t)\) is the smooth boundary displacement carried by the
characteristic flow. This is the local differential of Hamilton's
principal function evaluated on the transported endpoint perturbation,
not an imposed bulk-noise ansatz. Hence
\begin{equation}
\nabla S_\partial
=
(\nabla^2S)\eta .
\label{eq:Hessian_filtering}
\end{equation}
The Hessian of the principal action is therefore the first filter
converting boundary uncertainty into momentum uncertainty.

At the effective level we use the full material derivative
\(D_t=\partial_t+v\cdot\nabla\). At the order retained here, replacing
the reference transport \(D_t^{(0)}\) by the full characteristic
transport modifies only higher-order terms in \(S_\partial\). We
therefore write
\(\zeta=-D_tS_\partial\), giving
\begin{equation}
\xi
=
-\nabla D_tS_\partial .
\label{eq:force_from_material}
\end{equation}

For smooth colored boundary fluctuations, the non-commutation of
\(\nabla\) and \(D_t\) gives
\begin{equation}
\xi_i
=
-
S_{ik}\dot\eta_k
-
(D_tS_{ik})\eta_k
-
(\partial_i v_j)S_{jk}\eta_k ,
\quad
S_{ik}=\partial_i\partial_kS .
\label{eq:general_inherited_force}
\end{equation}
Equation~\eqref{eq:general_inherited_force} is the central result. The
inherited force contains phase-inertial, transported-curvature, and
convective-deformation sectors. In all cases, boundary fluctuations are
geometrically filtered by the Hamilton--Jacobi action field before
appearing as effective force noise. The commutator derivation and the
two-boundary generalization are given in the Supplemental Material.

\paragraph*{Stochastic limit and non-equivalence.}

A white-noise Langevin form is obtained only after coarse graining the
smooth boundary fluctuations over timescales longer than their correlation
time. In this short-correlation limit, the transported boundary imprint
may be represented in Stratonovich form \cite{Stratonovich1966,Gardiner2009},
\begin{equation}
dS
+
H(q,\nabla S,t)\,dt
=
\alpha(q,S,\nabla S,t)\circ dW_t ,
\label{eq:stochastic_HJ}
\end{equation}
which preserves the chain rule inherited from the smooth-noise limit.
Taking the gradient gives
\begin{equation}
dp_i
=
-\partial_iH\,dt
+
\partial_i\alpha_a(q,S,\nabla S,t)\circ dW_a .
\label{eq:stochastic_momentum}
\end{equation}

The endpoint origin constrains the noise amplitude. For a
boundary-response tensor \(\beta_{ak}(q,t)\), the natural coupling is
\begin{equation}
\alpha_a
=
\beta_{ak}(q,t)\partial_kS ,
\label{eq:boundary_coupling}
\end{equation}
so that
\begin{equation}
\partial_i\alpha_a
=
(\partial_i\beta_{ak})\partial_kS
+
\beta_{ak}S_{ik}.
\label{eq:gradient_alpha}
\end{equation}
For homogeneous boundary coupling, the effective diffusion tensor becomes
\begin{equation}
D_{ij}(q,t)
=
\frac12
\sum_a
\partial_i\alpha_a\,\partial_j\alpha_a
=
\frac12
\sum_a
\beta_{ak}\beta_{a\ell}S_{ik}S_{j\ell}
\propto
S_{ik}S_{jk},
\label{eq:Hessian_diffusion}
\end{equation}
where the final proportionality holds for isotropic homogeneous boundary
response.

This establishes the non-equivalence with homogeneous additive Langevin
noise. Although a phenomenological covariance can always be matched
locally, the inherited stochastic forcing cannot be represented globally
by constant additive noise unless the Hessian modulation is effectively
uniform over the explored region. The stochastic calculus limit and the
corresponding It\^o drift correction are given in the Supplemental
Material.

\paragraph*{Harmonic oscillator benchmark.}

For the harmonic potential \(V(q)=kq^2/2\), the principal function can be
written as
\begin{equation}
S(q,t)
=
\frac12A(t)q^2 ,
\label{eq:harmonic_action}
\end{equation}
with
\begin{equation}
\dot A
+
\frac{A^2}{m}
+
k
=
0 .
\label{eq:Riccati_main}
\end{equation}
For a smooth endpoint displacement \(\eta(t)\),
\(S_\partial=A(t)q\,\eta(t)\). Using \(\zeta=-D_tS_\partial\) and taking
the gradient gives
\begin{equation}
\xi(t)
=
k\,\eta(t)
-
A(t)\dot\eta(t).
\label{eq:oscillator_force_main}
\end{equation}
Thus even the harmonic oscillator does not inherit a generic additive
noise. The force separates into a positional sector, \(k\eta\), set by
the restoring force, and a phase-inertial sector, \(-A\dot\eta\), set by
the local curvature \(A(t)=\partial_q^2S\) of Hamilton's principal
function.

If the smooth boundary fluctuations have correlator
\(C_\eta(\tau)=\langle\eta(t)\eta(t+\tau)\rangle\), and \(A(t)\) varies
slowly over the boundary correlation time, the force correlator is
\begin{equation}
C_\xi(\tau)
=
k^2C_\eta(\tau)
-
A^2C_\eta''(\tau).
\label{eq:force_correlator_main}
\end{equation}
The inherited force is therefore generally colored even for
short-correlated boundary fluctuations; the white Langevin limit appears
only after Markovian coarse graining. Derivations and the Rayleigh-damped
extension are given in the Supplemental Material.

\paragraph{Overdamped limit.}
With an independent Rayleigh friction term, the effective dynamics becomes
\(m\ddot q+\gamma\dot q=-\nabla V+\xi(q,t)\), with \(\xi=\nabla\zeta\).
In the rapid momentum-relaxation regime,
\(m\ddot q\ll\gamma\dot q\), this reduces to
\[
\gamma\dot q=-\nabla V+\xi(q,t).
\]
Thus the boundary-induced force survives the overdamped projection while
retaining the Hessian filtering discussed above.

\paragraph*{Outlook.}

The same variational logic extends naturally to fields, where fluctuating
boundary data may live on spatial boundaries or Cauchy hypersurfaces. In
that case the field momentum
\(\Pi(x,t)=\delta S[\phi,t]/\delta\phi(x)\) acquires a stochastic force
density by functional differentiation of the transported boundary-action
source, in direct analogy with \(\xi=\nabla\zeta\). Thus field dynamics
inherits the geometry of the action functional just as particle dynamics
inherits the geometry of Hamilton's principal function. Details of this
field-theoretic extension and its relation to boundary-driven SPDEs are
given in the Supplemental Material.

\paragraph{Conclusion.}
We have shown that effective Langevin forcing can arise from fluctuating
variational boundary data rather than imposed bulk noise. The force is
the gradient of a transported perturbation of Hamilton's principal
function and is therefore constrained by the action-field geometry. This
yields multiplicative, state-dependent noise, with homogeneous additive
forcing recovered only after coarse graining. Fluctuating variational
boundaries therefore provide a variational route to stochastic dynamics
beyond phenomenological noise models.\\

\begin{acknowledgments}
The author thanks Jes\'us Fern\'andez-Castillo for fruitful discussions.
Support was provided by the Agencia Estatal de Investigación (AEI, Spain)
under Grants TED2021-132296B-C52 and CPP2024-011880, and by Fundación
BBVA--Programa Fundamentos 2025.
\end{acknowledgments}

\bibliographystyle{apsrev4-2}
\bibliography{bibliography}

\end{document}